# FORMATION OF GLOBAL TENDENCIES :
# SCIENTIFIC HYPOTHESIS ABOUT DESTRUCTION
# OF CIVILIZATIONS

Viktor I. Shapovalov (1), Nickolay V. Kazakov (2)

(1) The Volgograd Branch of Moscow Humanitarian-Economical Institute, Volgograd, Russia,
(2) The Volgograd State Technical University, Russia
shavi@rol.ru

The new explanation of global tendencies (in particular, of natural calamities and other disasters, taking place in present time in different countries) is suggested.

Probably, nobody doubts that for past years there have been done not best changes in our relations with nature. Let us emphasize the main idea: natural calamities, ecological crisis, technogenical catastrophes, military and social conflicts have the general property - they bring destruction i.e. increase medium entropy (disorder). But entropy change is regulated by the fundamental laws of nature. In other words, to our opinion, in present time the development of dangerous natural tendency is observed - the tendency of destruction prevailing.

It is possible to comprehend the reasons of this tendency due to exploring of the laws open rather recently and by virtue of it have not yet received wide popularity. These laws characterize the entropy change in the opened system [1-5]. In the concise form these regularities can be formulated as two proposals:

1. Each opened system has a special level of ordering named critical one[1]. If a system is ordered lower than critical level, then the processes of order increasing prevail; if ordering is higher than critical level, then the processes of disorder increasing prevail; at the very critical level the named processes are balanced and a system gets stable stationary state.

2. The value of critical level rigidly corresponds to the value of interaction of system with external environment, i.e. to openness degree of system. Thus if one wants to increase internal order of interesting system then one ought to increase its openness degree, the new value of which corresponds to new, higher critical level of ordering. As a result the processes of order increasing of system to new critical level will prevail. Vice versa, for disorganization one ought to decrease its openness degree. It will decrease critical level and cause prevailing of processes of decrease of order in system up to the new value of critical level.

So it is possible to change critical level of order only by changing openness degree of system.

Many events of our life illustrate the given legitimacies. To support itself in the good physical form, we are engaged with various power exercises, that is we subject ourselves to exterior action, causing inside an organism processes of ordering. The impairment of a mode of a physical training inevitably gives in some disorganization of our organism. Establishing rules impeding access at the inside market qualitative and (or) of the cheap goods of foreign production, thus we increment closure of that branch of a domestic economy, which yields the similar goods. As a result in it will inevitably arise some processes of disorganization, that, at the end, will be shown in deterioration of quality of the goods yielded by the given branch. Other examples can be found in [1-3,5,6].

---

[1] Critical level of ordering - level of system order corresponding to stable stationary state [3,4].

Now we shall explain the relation of told above to the global problems, to which mankind should recently collide even more often.

By its statistical expression entropy is connected to the probability of events. In practice the action of entropy laws change the probability of events. In particular, the events promoting realizations of these laws occur more often others.

While transforming surrounding world mankind changes its ordering. System "The Earth" as any unclosed system has its own critical level of organization. The value of critical level is rigidly correlated to the degree of openness that is to the degree of interacting of planet and space. According to the described above regularities if the system "The Earth" is ordered low critical level then processes of order increasing prevail on it; if order is above then critical level - processes of disorganization prevail. In the first case mankind transforming surrounding world in total increases more order than disorder in it. For how long time it will be? - To such times when during its creative activity mankind will exceed the critical level of planet ordering. In this case the disorganization processes will prevail and probability of destroying processes increases and the surplus of constructed by mankind exceeding the critical level will be terminated (or ought to compensate by demolitions at surrounding media). By inertia the destructions will be a little more then needed ones for reaching the critical level. Under critical level the processes of ordering will prevail and mankind will build houses, partition rivers and succeed in other creative activity, that is in Earth entropy decreasing. After some time mankind will again exceed critical level and processes obliterating surplus of mankind creations will prevail, and so on [5].

The system being above critical level initiates a wide spectrum of processes capable to destroy order surplus in it. It is easy to understand that in the case of the Earth there are wars among these fastest processes. Let's notice, that for two last centuries the sole processes effectively braking total volume of human construction are the world wars. In other words, while the mankind is engaged in peace time in transformation of a nature, it inevitably comes nearer to a critical level, and consequently, to war and (or) to such calamity of nature, which on scales and speed of destruction is comparable to war. It is possible to judge that on a planet the critical level of ordering is already exceeded by occurrence of the characteristic tendency: increasing of intensity of natural calamities, destroying climate change, aggravation of ecological crisis, appreciable increasing of probability of incidents and accidents, technogenical accidents, epidemics, social conflicts, local wars, and other events promoting the disorder. The well-known "hotbed effect", considered to be responsible for global warming, appears to be only a part of the specified tendency.

The rather constant openness degree of the Earth sets the constant critical level of ordering of the Earth. The mankind, creating while peace time, inevitably aspires to exceed this level. And when it occurs, then in spite of on any peace initiatives and ecological programs, the pro-cesses of disorganization at the Earth should have to prevail (i.e. should raise probability of any events resulting in destructions). At the same time increase of an openness of a planet (for example, the development of the Moon, then come a turn, for example, Mars and etc) would raise also the value of its critical level, that would result in prevalence of processes of ordering, and only then the ecological programs could effectively restore natural environment, and the mankind would come to a state of steady peace existence [4,6].

So, the sole way to increase a critical level of the system "Earth" (including mankind as a subsystem) is to increase its openness. Otherwise destructive tendency described here will inevitably be finished by global accident (i.e. world war or something comparable with it). The specified openness can be increased by inclusion in sphere of human industrial activity of the nearest proportional to the Earth space object - Moon. For this purpose it is necessary to unit efforts of the most advanced countries. It is necessary to do this as soon as possible, as it is clear that the tendency is present.

Unfortunately, to our view, there is a very small probability the governments of any countries will take into account the given caution. They do not know at all about the described here laws. In our opinion, much time can pass before these laws will be recognized even by global scientific community. Apparently, mankind will not have time to be rescued. The civilization will be lost.